\documentclass[aps,twocolumn,prl,showpacs,showkeys,floatfix]{revtex4}
\usepackage{epsfig}
\usepackage{isolatin1}
\usepackage{makeidx} \makeindex
\usepackage{amsmath} \usepackage{amssymb} \usepackage{amsthm}
\usepackage{mathrsfs}

\begin{document}

\title{Initial decoherence in solid state qubits}

\begin{abstract}
We study decoherence due to low frequency noise in Josephson qubits.
Non-Markovian classical noise due to switching
impurities determines inhomogeneous broadening of the 
signal. The theory is extended to include  effects of 
high-frequency quantum noise, due to 
impurities or to the electromagnetic 
environment. 
The interplay of slow noise with intrinsically 
non-gaussian noise sources may explain the rich physics 
observed in the spectroscopy
and in the dynamics of charge based devices.
\end{abstract}

\author{G. Falci} \affiliation{Dipartimento di Metodologie Fisiche e 
Chimiche (DMFCI), 
Universit\'a di Catania. Viale A. Doria 6, 95125 Catania (Italy) 
\& MATIS - Istituto Nazionale per la Fisica della Materia, Catania}
\author{A. D'Arrigo} \affiliation{Dipartimento di Metodologie Fisiche e 
Chimiche (DMFCI), Universit\'a di Catania. Viale A. Doria 6, 95125 Catania (Italy) 
\& MATIS - Istituto Nazionale per la Fisica della Materia, Catania}
\author{A. Mastellone} \affiliation{Dipartimento di Metodologie Fisiche e 
Chimiche (DMFCI), Universit\'a di Catania. Viale A. Doria 6, 95125 Catania (Italy) 
\& MATIS - Istituto Nazionale per la Fisica della Materia, Catania}
\author{E. Paladino} \affiliation{Dipartimento di Metodologie Fisiche e Chimiche (DMFCI), 
Universit\'a di Catania. Viale A. Doria 6, 95125 Catania (Italy) 
\& MATIS - Istituto Nazionale per la Fisica della Materia, Catania}

\email[email: ]{gfalci@dmfci.unict.it}


\maketitle

Considerable progress 
has been recently achieved
in implementing qubits with 
superconducting nanocircuits.
Coherent oscillations\cite{kn:Nakamura1,kn:vion,kn:Duty}
and
entanglement of coupled 
charge qubits\cite{kn:two-qubit} have been
observed.
Limitations in the performances arise from noise due
to material and device dependent
sources\cite{kn:rmp,DPG,PRL,kn:galperin,kn:makhlin,kn:averin04}.
Noise due to individual impurities behaving as bistable 
fluctuators (BF) is 
a severe source of dephasing for charge based devices.
Sets of BFs determine $1/f$ 
noise\cite{kn:weissman,kn:zorin}, and effects due to 
individual BFs 
has been observed both in spectroscopy and in time resolved 
dynamics\cite{Duty,Saclay}. 
Observations
show a variety of features, as the drastic reduction of 
the amplitude of the coherent signal\cite{kn:Nakamura1,kn:vion,Saclay}  
or relaxation limited decoherence\cite{kn:Duty},
strongly dependent on the particular device 
and on details of the protocol\cite{kn:vion,nakamura-echo,Varenna}. 
Theories BFs environments\cite{Varenna,PRL,DPG,kn:galperin,kn:bruder} 
allow to understand 
several physical aspects, although a quantitative 
framework embedding the variety of phenomena is still 
missing. 
Phenomenological models of the environment as a suitable 
set of harmonic 
oscillators\cite{kn:rmp,kn:makhlin,kn:averin04} have also
been studied. While they 
are unable to describe aspects related to the discrete 
nature of noise\cite{Varenna,PRL,DPG,kn:galperin,kn:bruder},
gaussian environments may 
sometimes provide useful information.

In this work we study numerically a model of 
discrete noise which potentially 
explain the experimental features due to $1/f$ noise, 
and seek a classification of the possible effects on 
the basis
of simple theoretical arguments. In particular we study 
inhomogeneous broadening due to slow noise and its interplay 
with additional noise sources, pointing out
that the presence of BFs may pose reliability
problems for charge based devices.

We consider a qubit 
anisotropically\cite{kn:rmp} 
coupled to classical stochastic process 
$\xi(t)$. The Hamiltonian is 
\begin{equation}
\label{eq:hamiltonian}
H \,=\,  H_Q \,-\, {1 \over 2}\, \xi(t) \, \sigma_z
\end{equation}
where 
$H_Q = - {1 \over 2} \vec{\Omega} \cdot \vec{\sigma}$
refers to the qubit. Both the operating point, i.e. the angle 
$\theta$ between $\hat{z}$ and $\vec{\Omega}$, and 
the splitting $\Omega$ are tunable. This also modulates 
sensitivity to noise. For weak coupling  
the relaxation
$T_1^{-1} = s^2  \, S(\Omega)/2$ and the dephasing rate
$T_2^{-1} = (2 T_1)^{-1} + T_2^{\prime \,\,-1}$,
$T_2^{\prime \,\, -1} =  c^2\, S(0)/2$ being the 
adiabatic term which gives secular 
broadening\cite{kn:slichter}, 
are tuned by $c=\cos\theta$ and 
$s=\sin\theta$.
Only the power spectrum of noise,
$S(\omega)= \langle \xi \xi \rangle_\omega$, enters
therefore in weak coupling the qubit is sensitive only to 
properties of the environment at the level
of two point correlations. This picture breaks down if
the environment extends to low frequencies\cite{kn:cohen}.
For instance
Random Telegraph Noise (RTN) due to a single BF,
$\xi(t)=\{0,v_0\}$, switching at a rate $\gamma_0$ 
is slow if $g_0=(\Omega^\prime-\Omega) /\gamma_0 >1$~\cite{DPG},
 where the qubit frequencies 
$\Omega$ and
$\Omega^\prime=\Omega [(v_0/\Omega+ c)^2 + s^2]^{1/2}$ 
correspond to the two values of $\xi$. 
This model describes an incoherently switching 
charged impurity close to a qubit.
For $g_0>1$ features 
of the discrete nature of the BF
become apparent\cite{DPG}.

A set of $N_{bf}$ BFs ($\xi_i$) switching
at rates $\gamma_i$, coupled with the qubit via 
$\xi(t) = \sum_i \xi_i(t)$, models $1/f$ noise if  
$\gamma_i$ are distributed\cite{kn:weissman} with 
$P(\gamma) \propto 1/ \gamma$. 
The BF-$1/f$ spectrum is
$S^{1/f}(\omega)=  \sum_i \, \frac{1}{2} 
\, v_i^2 \,\gamma_i/(\gamma_i^2 + \omega^2)$, and 
if $\gamma_i \in [\gamma_m,\gamma_M]$, in the
same interval of frequencies is approximated by
$S^{1/f}(\omega) \approx
 [(\pi/4) 
\,N_{bf} \overline{v^2} / 
\ln (\gamma_M/\gamma_m)]\,\,\omega^{-1}$. 
Noise extends for several decades and in particular
slow BFs ($g_i<1$), an environment with memory, make 
unstable the calibration of 
the device. Hence the qubit dynamics will 
depend on details of the protocol.
Decoherence due to BFs $1/f$ noise 
for various protocols  
has been studied for $\theta=0$,
where exact solutions are available\cite{PRL}.
On the other hand the splitting is less sensitive to 
fluctuations at
optimal working point\cite{kn:vion}, 
${\theta}=\pi/2$ (parameters $g_i$ become smaller),
and part of the effects of the slow noise is eliminated 
(at lowest order $T_2^{\prime \,\, -1}$ vanishes).

Ideal quantum 
protocols assume measurements of 
individual members of an ensemble of identical 
(meaning that preparation is controlled) evolutions 
of the qubit, defocusing occurring only {\em during} 
the time evolution. In 
practice for solid-state devices one collects several 
qubit evolutions, in 
an overall measurement time $t_m$. Lack of control 
on the environment preparation
determines defocusing of the 
signal, 
analogous to inhomogeneous 
broadening in NMR\cite{kn:slichter}. This is also true for 
single-shot measurements\cite{kn:nec-single-shot}. In our case BFs active in additional broadening have 
$\gamma_i>{\gamma}^*\sim \min\{\overline{v}/10,t_m^{-1}\}$\cite{Varenna}.

We first study Hamiltonian 
(\ref{eq:hamiltonian})
by simulating the Stochastic Schr{\"o}dinger 
Equation for the qubit
in a BF-$1/f$ environment.
We generate $\xi(t)$ as a sum of 
$N_{bf} \le 2000$ RTN processes with proper
distribution of parameters. In order to minimize errors in
generating $\xi(t)$ we use a ``waiting time'' 
algorithm~\cite{kn:petruccione}, which also reduces 
the computational time.
The qubit 
propagator is evaluated as the product 
of the propagators between successive switches. 
Finally we perform the statistical average. 
We study an ensemble of 
time evolutions of the qubit, each lasting 
for a time 
$t$. During the overall time $t_m$ of the protocol the 
environment evolves in an uncontrolled way, so BFs with
$\gamma_i \gg 1/t_m$ average, whereas BFs with
$\gamma_i \ll 1/t_m$ are frozen. Thus for the simulation
we consider  
$\ge 10^5$ realizations of $\xi(t^\prime)$, for 
$0<t^\prime<t$. For the individual BFs at 
$t^\prime=0$ we choose {\em the same} 
initial $\xi_i(0)=0,1$ 
if $\gamma_i < 1/t_m$ whereas if  $\gamma_i > 1/t_m$
we take a distribution with $0 <\overline{\xi_i(0)} <1$.
This prescription has been checked against more 
accurate ones in Ref.\cite{Varenna}. 
\begin{figure}[t]
\centering
\epsfxsize=3.2in\epsfysize=1.85in\epsfbox{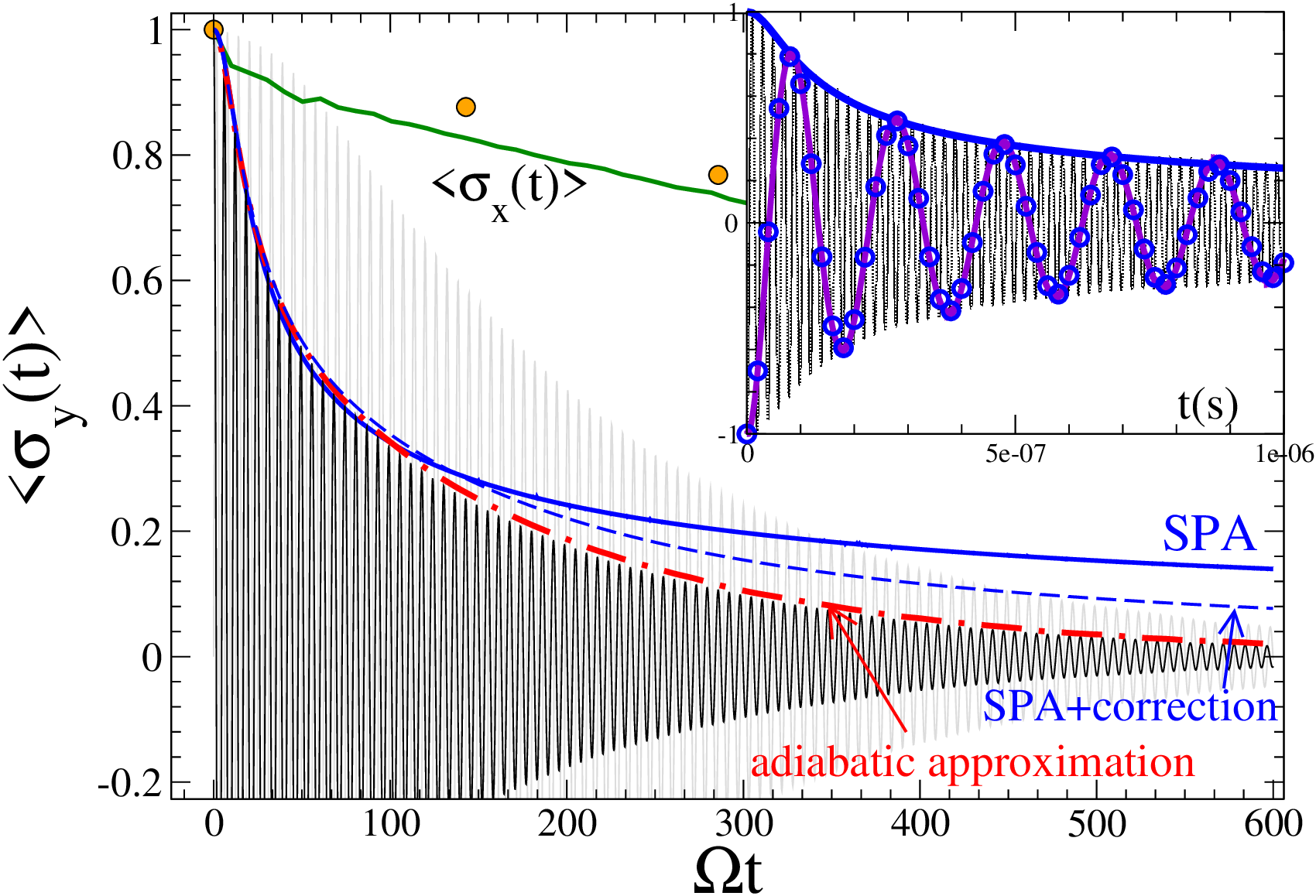}
\caption{Simulations of an adiabatic 
$BF-1/f$ environment at $\theta=\pi/2$.
Relaxation studied via $\langle \sigma_x \rangle$ (green
line) is well approximated by the weak coupling 
theory (dots). Dephasing in repeated measurement damps the
oscillations (thin black line). Part of the signal 
is recovered if the environment is recalibrated (thin gray 
line). 
Noise is 
produced by $n_d=250$ BFs per decade, 
with $1/t_m=10^5$ rad/s $\le \gamma_i \le \gamma_M = 
10^9$ rad/s $< \Omega = 10^{10}$ rad/s. The coupling
$\bar{v} = 0.02 \,\Omega$ 
is appropriate to charge 
devices, and corresponds to 
$S(\omega)= 16 \pi  A E_C^2/\omega$ with $A \sim 10^{-6}$~\cite{kn:zorin}. 
The adiabatic approximation Eq.(\ref{eq:blur}) fully
accounts for dephasing (red dot-dashed line). 
The Static Path Approximation 
(SPA) Eq.(\ref{ref:quadratic}) (blue solid line) and 
the first correction (blue dashed line) 
account for the initial suppression, and it is valid
also for times $t \gg 1/\gamma_M$. 
In the inset Ramsey fringes with parameters appropriate
to the experiment~\cite{kn:vion} (thin black lines).
The SPA (blue solid line), 
Eq.(\ref{ref:quadratic}), is in excellent agreement
with observations~\cite{Saclay},  
and also predicts the correct phase shift of the 
Ramsey signal 
(blue dots, compared with simulations for small 
detuning $\delta=5\,Mhz$, violet line), which tends to
$\approx \pi/4$ for large times.} 
\label{fig:1}
\end{figure}

Results at $\theta=\pi/2$ for an adiabatic $1/f$ environment, 
$\gamma_M  \ll \Omega$, show the presence of several 
time scales (Fig.~\ref{fig:1}). Coherent
oscillations of $\langle \sigma_y \rangle$ are 
initially suppressed with a power law.
Relaxation occurs on much longer time scales, given by the 
weak coupling result. The initial suppression is 
due to inhomogeneous broadening. This is apparent if 
we compare with
results with a feedback protocol
simulated by resetting $\xi(0)$ at the same value for 
each realization of $\xi(t^\prime)$. 

\begin{figure}[t]
\centerline{
\epsfxsize=3.2in\epsfysize=1.510in\epsfbox{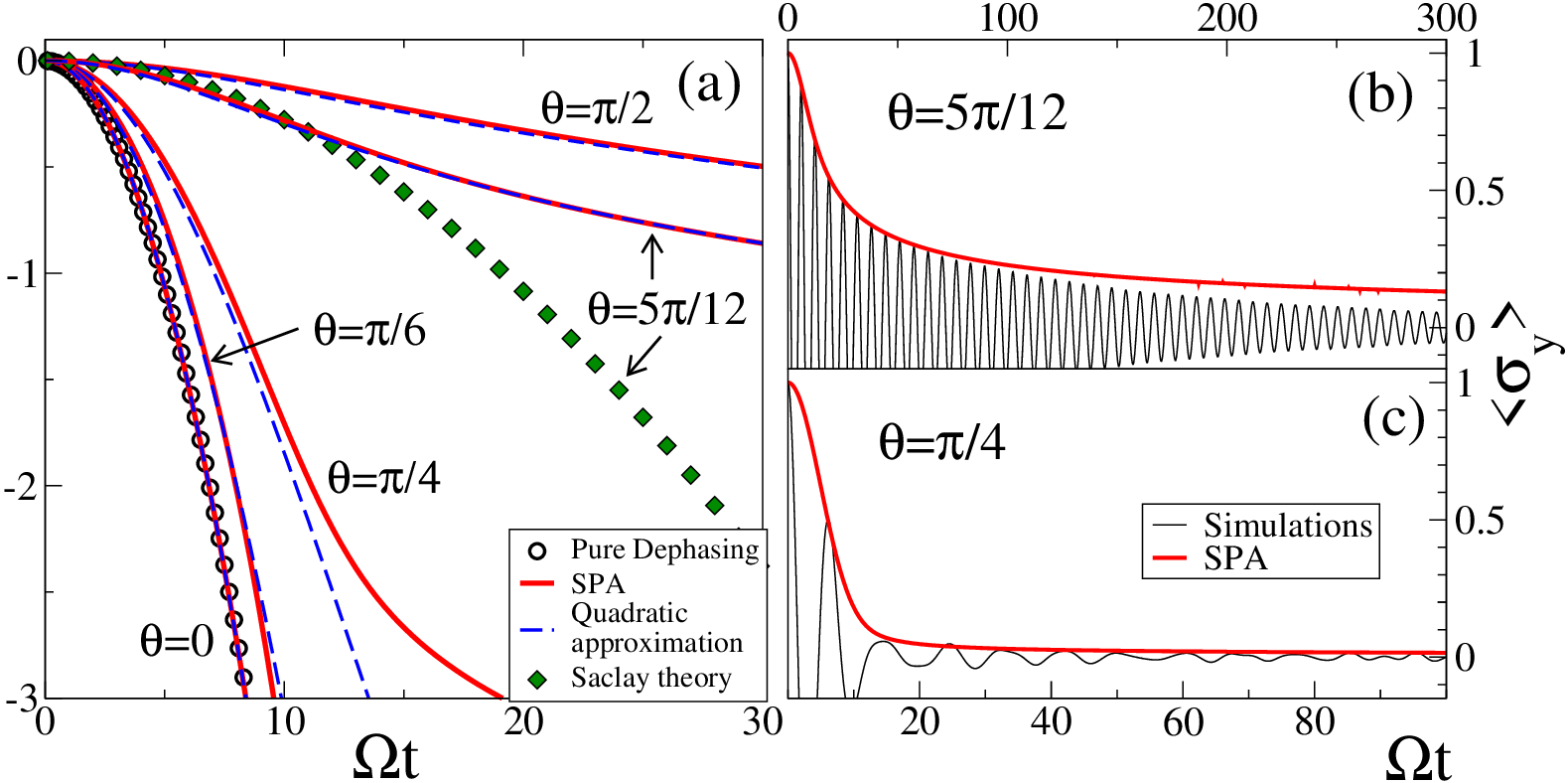}
}
\caption{Signal decay at non optimal points. 
(a) The SPA (solid red lines) is compared with the 
quadratic approximation 
Eq.(\ref{ref:quadratic}) (dashed blue lines) at different bias 
points. Eq.(\ref{ref:quadratic})
works near $\theta=0$ and $\theta=\pi/2$. 
The exact result at $\theta=0$ is also indicated 
(dots). This result multiplied by $\cos^2 \theta$, 
is often used also at $\theta\neq0$ for interpreting 
experiments (``Saclay theory'', diamonds).
Out of the optimal point the SPA agrees with 
results of the simulations
with BF-$1/f$ noise, 
for the set of Fig.\,1, at $\theta=5\pi/12$ (b) and 
$\theta=\pi/4$ (c).
}
\label{fig:2}
\end{figure}

Negligible relaxation allows to treat $\xi(t)$ 
in the adiabatic approximation.
Observables are then given by path-integrals over
a weight $P[\xi(t)]$ of the stochastic process. 
We study 
the  averaged phase shift $\Phi(t)$, defined 
as
\begin{equation}
\label{eq:blur}
\mathrm{e}^{- i  \,\Omega t - i \,\Phi(t)} \,=\, 
\int\!\! 
{\cal D} \xi(t^\prime) \; P[\xi(t^\prime)] \; 
\mathrm{e}^{- i \int_0^t dt^\prime  \, \Omega(\xi(t^\prime))}
\;
\end{equation}
which gives the decay of the qubit coherences, 
$\langle \sigma_y\rangle \propto 
\mathrm{exp}{[\Im \Phi(t)]}$. 
Here $\Omega(\xi(t)) = 
\Omega [(\xi(t)/\Omega+ c)^2 + s^2]^{1/2}$ is the 
instantaneous qubit splitting. Numerical 
evaluation of 
the path-integral Eq.(\ref{eq:blur}) fully agrees with
the simulations. 
Further insight is obtained by approximating 
Eq.~(\ref{eq:blur}). 
The Static-Path Approximation (SPA), 
$\xi(t^\prime)=\xi_0$ 
accounts for lack of control on the environment 
preparation via a statistically distributed $\xi_0$.   
This blurs of the overall signal, 
an effect analogous to the 
rigid lattice line breadth in 
NMR~\cite{kn:slichter}.
For a set of BFs, 
if $N_{bf}$ is large enough $\xi_0$ is gaussian 
distributed with variance 
$\sigma^2_{\!\xi} = \overline{v^2}N_{bf}/4 = 
\int_0^\infty \, 
(d \omega/\pi) \, S(\omega)$, where it is intended that 
we consider only active BFs, $\gamma_i > \gamma^*$. 
The result, plotted in Figs.\ref{fig:1},\ref{fig:2} 
accounts for the initial suppression of the signal, 
showing that this latter is entirely due to inhomogeneous 
broadening.
Therefore the analysis of the 
initial 
suppression may give information on the the amplitude of 
$1/f$ noise at intermediate 
frequencies $1/t_m < \nu < 1/t$.
A good estimate of the SPA is obtained by a quadratic 
expansion of $\Omega(\xi_0)$ in $\xi_0$ (see Fig~\ref{fig:2}a)
\begin{equation}
\label{ref:quadratic}
- i \, \Phi(t) = - {1 \over 2} \,
{(c \, \sigma_{\!\xi} t)^2 \over 1 +  i 
s^2 \sigma^2_{\!\xi} t/\Omega } - {1 \over 2} \,
\ln \Big(1 +  i 
s^2 {\sigma^2_{\!\xi} t \over \Omega} \Big)
\end{equation}
which is accurate close to $\theta=0$ and $\theta=\pi/2$ for 
$\sigma_{\!\xi}/ \Omega$ small enough.
The 
resulting suppression factor
$\exp (\Im \Phi)$ turns from a 
$\, \exp (- {1 \over 2} c^2 \sigma^2_{\!\xi} t^2)$ behavior
at $\theta \approx 0$ to a power law,  
$[1 + (s^2 \sigma^2_{\!\xi} t/\Omega)^2 ]^{-1/4}$, 
at $\theta \approx \pi/2$. These limits  
reproduce known results for gaussian $1/f$ environments, 
namely at $\theta = 0$
the $t \ll 1/\gamma_M$ limit of the exact result~\cite{kn:palma} 
and for $\theta = \pi/2$ the short-time result of the 
diagrammatic approach of Ref.~\cite{kn:makhlin}.
This is not surprising since the SPA 
does not require knowledge about the {\em dynamics} 
of the noise sources, provided they are slow\cite{kn:nota}. 

Eq.~(\ref{eq:blur}) can be systematically 
approximated by sampling better 
$\xi(t^\prime)$ in $[0,t]$. For the first
correction $P[\xi(t)]$ is approximated by 
the joint distribution
$P(\xi_t t ; \xi_0 0)$, where 
$\xi_t = \xi(t)$.
At $\theta=\pi/2$ for 
generic gaussian noise we find 
$$
i  \Phi(t) =  {1 \over 2} 
\ln \Big[1 +  i 
{\sigma^2_{\!\xi} [1 - \pi(t)] t \over \Omega} \Big]
+ {1 \over 2} 
\ln \Big[1 +  i 
{\sigma^2_{\!\xi} \pi(t)\,t \over 3 \, \Omega} \Big]
$$
where 
$
\pi(t) = {1 \over 2 \sigma^2} \int_0^\infty \, 
(d \omega/\pi) \, S(\omega) (1- \mathrm{e}^{-i \omega t})$
is a transition probability, depending on the stochastic process.
For Ornstein-Uhlembeck processes it reduces to the result
of Ref.~\cite{kn:averin04}. 
The first correction suggests that the SPA, in 
principle valid for $t <  1/\gamma_M$, may have   
a broader validity (See Fig.\ref{fig:1}). 
For $1/f$ noise due to a set of BFs
it is valid also for $t \gg 1/\gamma_M$,    
if $\gamma_M \lesssim \Omega$. Of course 
the adiabatic 
approximation is tenable if $t < T_1 
= 2/S(\Omega)$.

\begin{figure}[t]
\centerline{
\epsfxsize=3.0in\epsfysize=1.6in\epsfbox{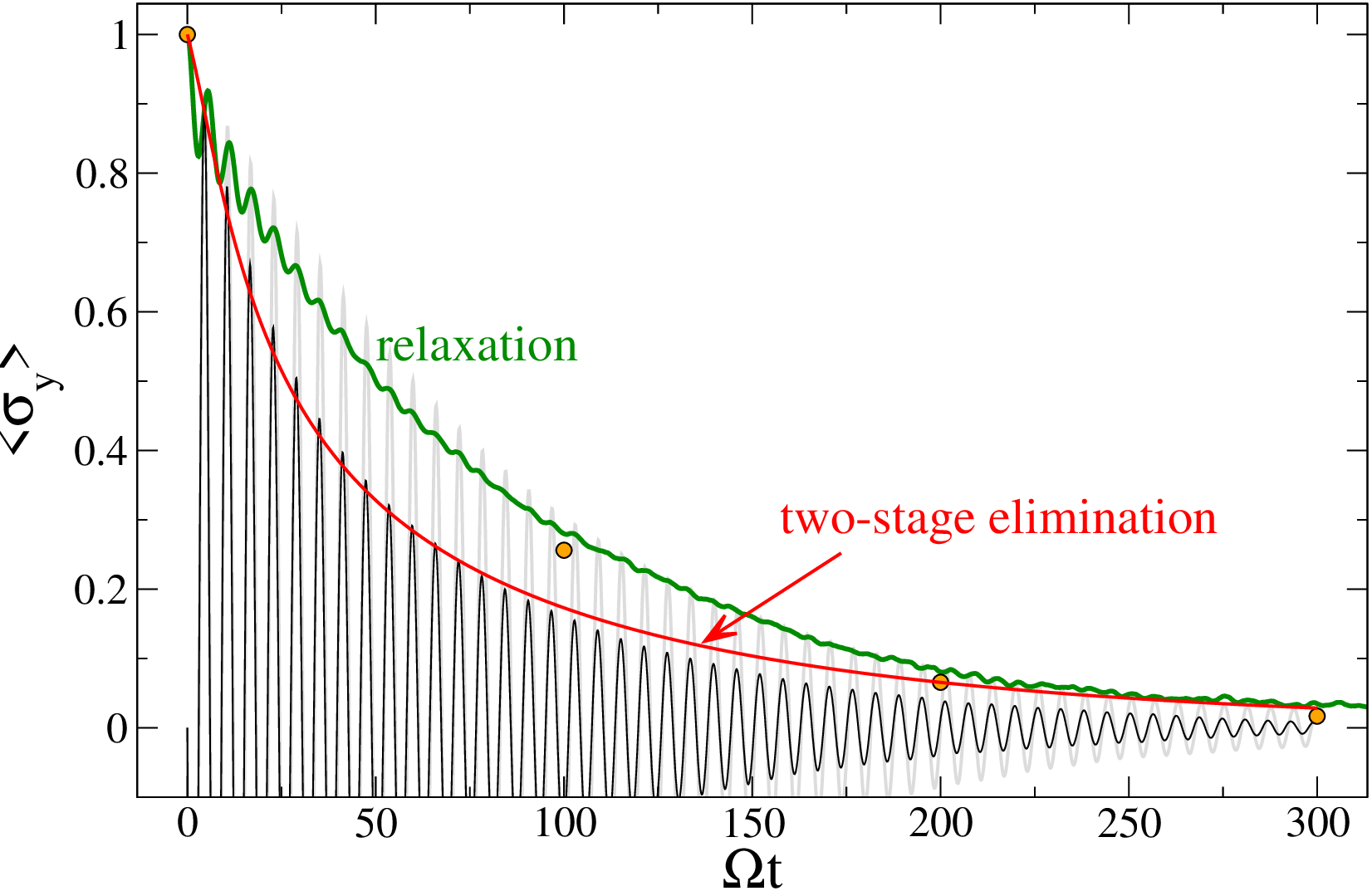}
}
\caption{Results of simulations with an adiabatic plus fast
$BF-1/f$ environment (same parameters of Fig.\,1 except for 
$\gamma_M = 10^{11}$ rad s$^{-1}$). Relaxation 
(thick solid green line) is given by the 
weak coupling result (dots). The initial suppression of the 
oscillation amplitude is partially removed by a feedback protocol
(shaded curves) and is well described by the two-stage 
elimination SPA theory (solid red line), Eq.(\ref{eq:slow-fast}).}
\label{fig:3}
\end{figure}

The main effect of faster BFs in the $1/f$ spectrum 
is the decrease of $T_1$. Relaxation is due only to the fast 
part of the spectrum $\omega \sim \Omega$ and well reproduced
by the Golden Rule. This is not true for decoherence, for 
instance in our example (Fig.~\ref{fig:3}) 
$T_2 \approx T_1$, 
as observed in NEC experiments~\cite{kn:Duty}. 
We study the interplay of fast and slow noise by  
a two-stage elimination. We first decompose 
$\xi(t) \to \xi(t) + \xi_f(t)$. 
Here 
$\xi(t)$ represents all BFs having switching rates small 
enough 
to be treated in the adiabatic approximation, 
$\gamma_i < \gamma_{ad}$ (in  
practice we may take $\gamma_{ad} \sim \Omega/10$).
Fast BFs are 
described by $\xi_f(t)$ or better modeled by a set of quantum
impurities as in Ref.~\cite{PRL}. In this case 
$\xi_f(t) \to \hat{\xi}_f$ and we have a 
quantum environment able to produce
also spontaneous decay. The 
reduced density matrix of the qubit can be written as
$$
\rho(t) \,=\, \int\!\! 
{\cal D} \xi(t) \; P[\xi(t)] \;\; \rho_f[t|\xi(t)]
$$
where $\rho_f[t|\xi(t)]$ is the qubit density matrix resulting 
from the elimination of 
the fast environment, under the ``drive'' $\xi(t)$, 
and can be found within the weak coupling theory~\cite{kn:nota2}. 
This is very simple if we treat slow noise in the SPA, 
where $\xi(t) = \xi_0$.
We are left with averages over $P(\xi_0)$ of the 
entries of $\rho_f[t|\xi_0]$. For instance 
the decay of the coherences
at $\theta=\pi/2$ is given by
\begin{equation}
\label{eq:slow-fast}
\mathrm{e}^{- {1 \over 4} \, S_f\!(\Omega) \, t  - {1 \over 2} \,
\ln \big| 1 +    
 \, \big[i \Omega + S_f\!(0)- {1 \over 2} S_f\!(\Omega) \big] 
\,  {\sigma^2_{\!\xi} t / \Omega^2}
\big| }
\end{equation}
where $S_f\!(\omega)$ refers to the set of fast BFs,
whereas $\sigma^2_{\!\xi}$ refers to the set slow BFs.
Eq.(\ref{eq:slow-fast}) agrees very well with simulations
(Fig.\ref{fig:3}). 

We notice that the validity of Eq.(\ref{eq:slow-fast}) is
not limited to 
the $\sim 1/\gamma$ distribution of 
switching rates
giving rise to $1/f$ noise.
According to this description relaxation 
and inhomogeneous broadening are due to separate sets of 
BFs. Therefore no special 
relation is expected to hold between $T_1$ and $T_2$.
The mixed term in Eq.(\ref{eq:slow-fast}), due to the 
interplay between slow and fast 
BFs does not qualitatively change this conclusion. Finally 
Eq.(\ref{eq:slow-fast}) is rather independent from the 
nature 
of the noise sources and the form of the spectrum and 
it is applicable in other situations, e.g. when  slow 
impurity noise combines with fast electromagnetic noise. 
Eq.(\ref{eq:slow-fast}) becomes exact if $\xi_f$ determines 
white noise, a scenario recently proposed to fit decoherence 
in phase-charge qubits.

We come now to effects of the discrete nature of 
noise. Results presented so far rely on the SPA and on the 
weak coupling theory, therefore they apply to situations 
where discrete and gaussian noise are indistinguishable.
Striking differences appear when only 
decoherence during time evolution~\cite{PRL,Varenna} matters,
or if the distribution of environment couplings $v_i$ is 
wide~\cite{kn:galperin}, the realistic scenario for the 
solid state. We now study the interplay of $1/f$ noise with 
RTN produced by one BF which is more strongly coupled with
the qubit. The model for the BF is minimal: it is an 
incoherent slow fluctuator, having  
$\gamma_0 \ll 1/t \ll \Omega$ but
$v_0 \lesssim \Omega$. Even if the BF is not 
resonant with the qubit\cite{kn:martinis} it strongly 
affects the output signal. 
If $g_0>1$ ~\cite{DPG}, it determines beats in the 
coherent oscillations and split peaks in spectroscopy, which
are signatures of a discrete environment. 
The additional BF makes bistable the 
working point of the qubit and amplifies defocusing due to 
$1/f$ noise. Even if the device is initially 
optimally polarized, during $t_m$ the BF may switch it 
to a different working point. %
The line shape of the signal will show 
two peaks, split by $\sim \Omega^\prime - \Omega$
and differently broadened 
by the $1/f$ noise in background. 
The corresponding 
time evolution will show damped beats, this phenomenology 
being entirely due to the non-gaussian nature of the 	
environment. For illustrative purposes we show results 
of a simulation at the optimal point, where $1/f$ noise 
is adiabatic and weaker than the typical noise level in 
charge qubits. This picture applies to smaller devices
The fact that even a single impurity on a $1/f$ 
background causes
a substantial suppression of the signal poses the problem 
of reliability of charge based devices. An  analytic
two-stage elimination combining
the SPA with the solutions for the dynamics of 
a qubit coupled to an impurity~\cite{PRL,DPG} can be developed, 
and will be 
presented elsewhere.
\begin{figure}[t]
\epsfxsize=1.8in \epsfysize=1.20in \epsfbox{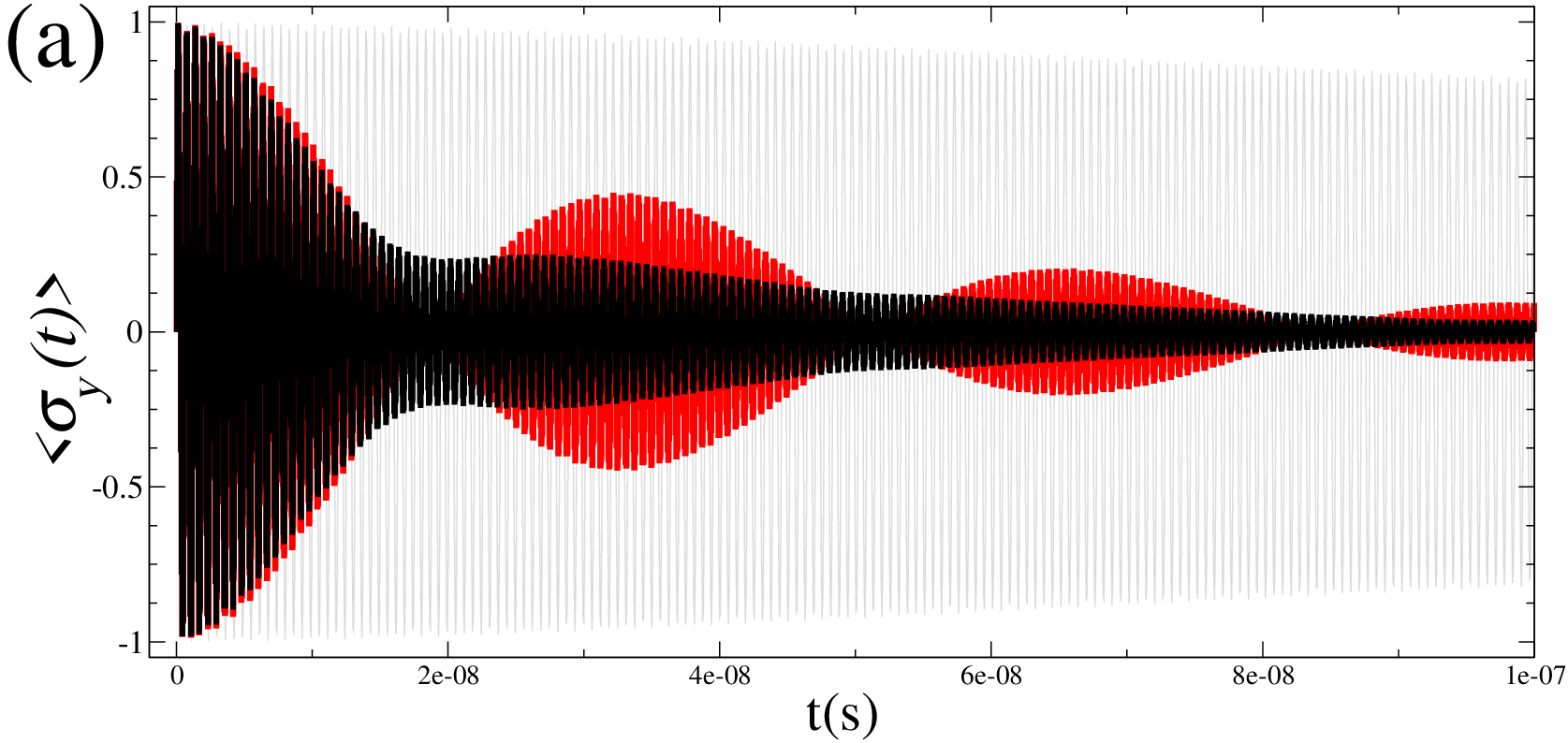}
\hfill
\epsfxsize=1.4in \epsfysize=1.23in 
\epsfbox{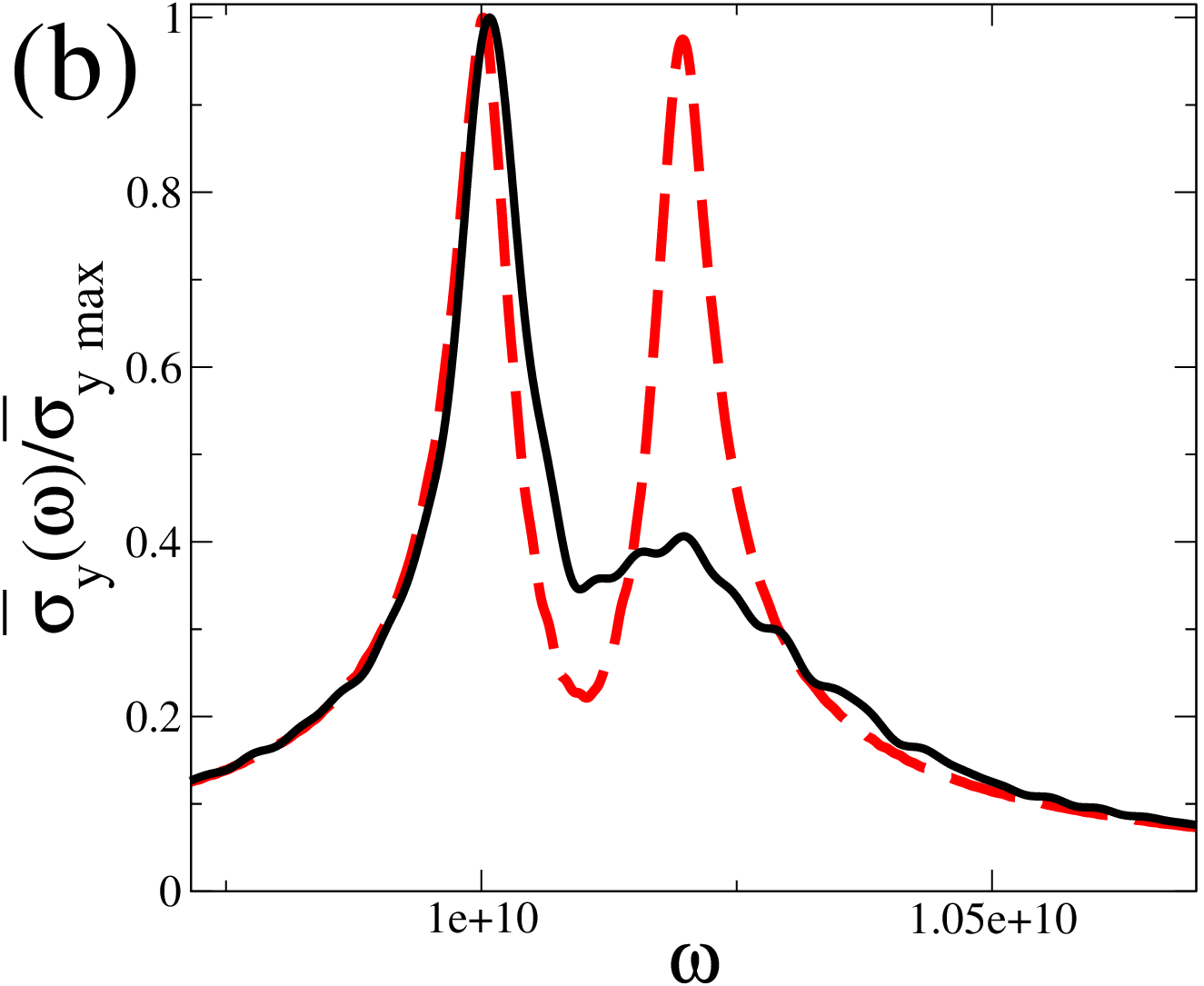}
\caption{ (a)
$\langle \sigma_y \rangle$ at $\theta=\pi/2$, $\Omega=10^{10}$Hz. 
The effect of weak adiabatic $1/f$ noise (light gray line)
($\gamma \in [10^5, 10^9]$ Hz, uniform $v=0.002 \,\Omega$,  
$n_d=250$) is strongly enhanced by adding a single slow 
($\gamma/\Omega=0.005$)
more strongly coupled ($v_0/\Omega=0.2$) BF 
(black line), which alone would give rise to beats 
(red line). (b) When the BF is present 
the Fourier transform of the signal may show a split-peak
structure. Even if peaks are symmetric for 
the single BF alone (dashed line), $1/f$ noise broadens them
in a different way (solid line).} 
 \label{fig:ft}
\end{figure}

Recently effects of the resonant coupling of the qubit with
a quantum two-level system, simulating defects in the 
tunnel oxide, have been proposed to explain 
features of the dynamics of phase  
Josephson qubits~\cite{kn:martinis}. We have shown that 
these effects are present even if the impurity behaves 
as a slow stochastic fluctuator. Our model describes
a very common situation in the solid 
state\cite{kn:galperin}, and it
is a minimal model for charge noise in 
charge and charge-phase qubits. Finally the interplay
between slow noise and fast noise whith 
generic spectrum is likely
to be important in general and can be studied with 
Eq.(\ref{eq:slow-fast}).
The main open 
questions are the accurate characterization 
beyond phenomenology of the physics 
of the noise sources and the design of specific strategies 
to defeat them and to improve reliability of devices. 

We acknowledge discussion with D. Esteve, R. Fazio, G. Ithier,
Y. Nakamura, G. Sch\"on and A. Shnirman. We acknowledge 
support from projects EU-SQUBIT2 (IST-2001-390083) and
MIUR-FIRB (RBAU01A9PM).


\begin{thebibliography}{0}
\bibitem{kn:nota} Eq.(\ref{ref:quadratic}) is also valid for 
Ornstein-Uhlembeck processes, see Ref.\cite{kn:averin04}, 
whereas for RTN the discrete nature of $\xi(t)$ modifies 
this result\cite{DPG}.
\bibitem{kn:nota2} Unless they are 
individually strongly coupled~\cite{PRL}, 
$\gamma_{ad} < v \sim \Omega$.
\end{thebibliography}
\end{document}